\documentclass[twocolumn,showpacs,superscriptaddress,10pt]{revtex4-1} 
\usepackage{amsmath,amssymb,graphicx,bm,color}

\RequirePackage[normalem]{ulem} 
\RequirePackage{color}\definecolor{RED}{rgb}{1,0,0}\definecolor{BLUE}{rgb}{0,0,1} 

\newcommand{\sgn}{\operatorname{sgn}}
\newcommand{\rect}{\operatorname{rect}}

\def\rev#1{#1}

\begin{document}

\title{Modifying the optical path in a nonlinear double-slit experiment}

\author{Vassilis Paltoglou}
\affiliation{Department of Mathematics and Applied Mathematics, University of Crete, 70013 Heraklion, Crete, Greece}
\author{Nikolaos K. Efremidis}
\email[Corresponding author: ]{nefrem@uoc.gr}
\affiliation{Department of Mathematics and Applied Mathematics, University of Crete, 70013 Heraklion, Crete, Greece}


\begin{abstract}
\rev{In this letter, we study a nonlinear interferometric setup based on diffraction rather than beam combining.} \rev{It consists of a} nonlinear analogue of Young's double-slit experiment where a nonlinear material is placed exactly after one of the slits. The presence of nonlinearity breaks the transverse spatial symmetry of the system and thus modifies the optical path. For moderate nonlinearities this leads to a self-induced shift of the intensity pattern in the transverse plane. A simple theoretical model is developed which is surprisingly accurate in predicting the intensity profile of the main lobes for a wide range of parameters. We discuss about possible applications of our model in nonlinear interferometry, \rev{for example in measuring the nonlinearities of optical materials}. 
\end{abstract}


\maketitle

The double-slit experiment is perhaps one of the most fundamental in quantum mechanics illustrating the wave-particle duality of a quantum wavepacket~\cite{feynman-vol3-1963}. The setup was originally introduced by Young to demonstrate the classical wave nature of light~\cite{young-ptrs1804}. Experiments were originally carried out with photons~\cite{taylo-pcps1909} until 1961 when the first experiment was performed using electrons~\cite{johns-zp1961}. 
In optics, the double-slit setup has been studied in the case of subwavelength plasmon plates~\cite{schou-prl2005}. Nonlinear extensions have been considered, in the case of self-focusing~\cite{roman-oe2006} as well as self-defocusing nonlinearities~\cite{sun-ls2008} where a double-slit aperture is used in front of a nonlinear material. In addition, such a configuration was used in connection to nonlinear Raman micorscopy~\cite{gache-prl2010}. 

\rev{The double-slit configuration is a main example of wave interference that is accomplished through diffraction. The dynamics of the optical wave which is mapped at the observation plane is affected by the intervening components. However, up to now, most interferometric setups do not utilize diffraction but rather direct beam combining resulting to discrete measurements. Typical examples of nonlinear interferometers were used for optical gating~\cite{dugua-apl1969} and to produce intensity-dependent phase shifts~\cite{milbu-prl1989} for optical logic gates. Interferometric setups can also be implemented with the use of optical fibers~\cite{agrawal-applications}.
}

\begin{figure}[tb]
\centerline{\includegraphics[width=\columnwidth]{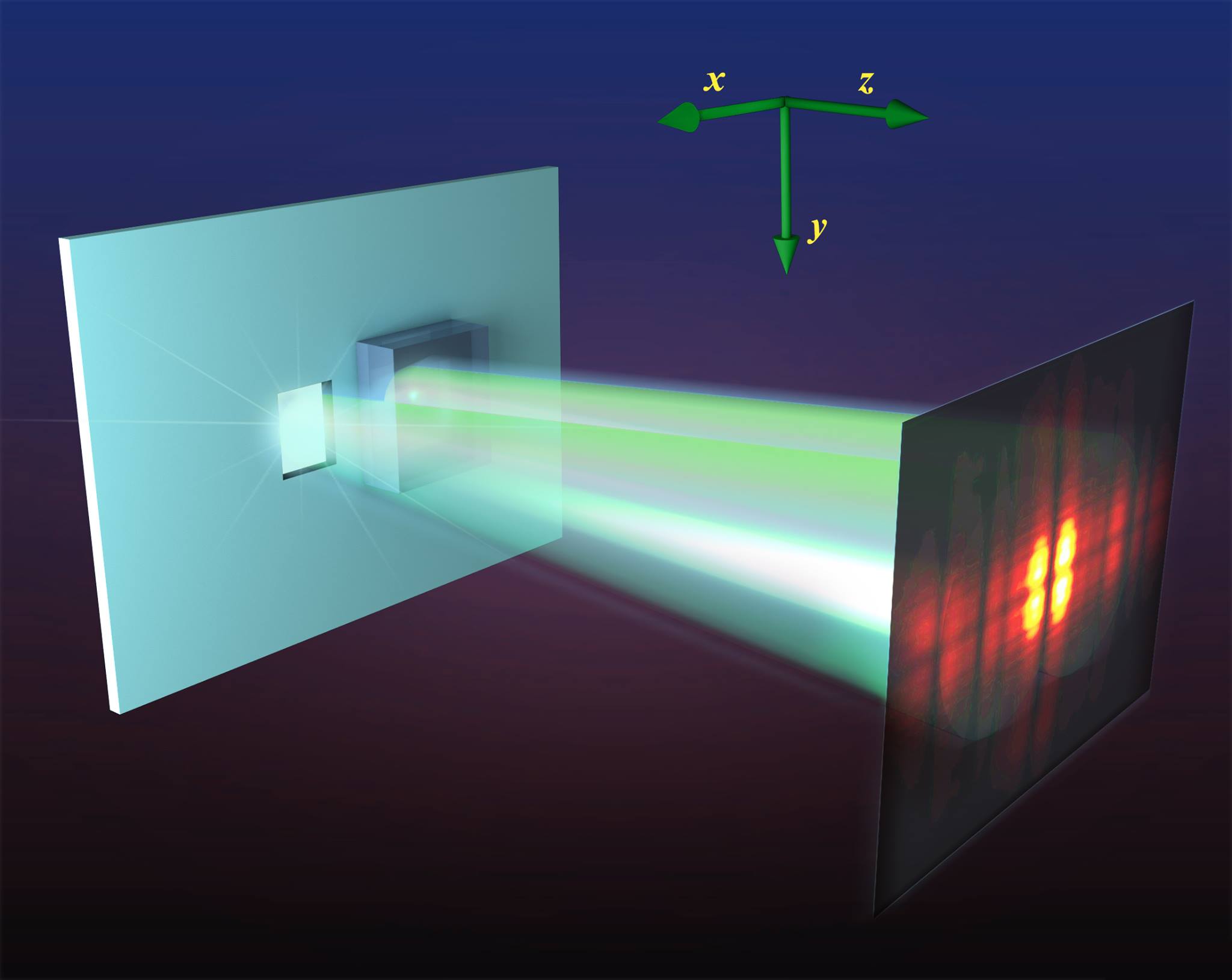}}
\caption{
(color online) A double-slit configuration with a nonlinear slab material placed in front of the left slit. The diffraction pattern is recorded at the output. \label{fig:1}}
\end{figure}

\rev{In this letter, we study a nonlinear interferometric setup based on diffraction rather than beam combining.}
\rev{It consists of a} nonlinear analogue of the double-slit experiment where a relatively thin slab of a nonlinear optical material is placed exactly after one of the slits and completely covers it (see Fig.~\ref{fig:1}). The presence of the nonlinear material modifies the optical path resulting to self-induced changes of the intensity pattern at the observation plane. In the case of weak to moderate nonlinearities the output intensity is spatially translated \rev{at the observation plane}. We show that the optical wave at the output can be accurately predicted by using a simple theoretical model. In the case of strong nonlinearity the system starts to behave in a more complicated fashion especially in the case of self-focusing nonlinearities. We expect that our results might be useful in nonlinear interferometry, for example, in measuring the nonlinear properties of optical materials~\cite{sheik-ol1989}.

Let us consider the double-slit configuration shown in Fig.~\ref{fig:1}. A coherent monochromatic laser light source with intensity $I_0$ is normally incident at the aperture plane ($z=0$). The two rectangular slits have dimensions $w_x\times w_y$ and their centers are separated by a distance $w$ along the $x$-direction. A nonlinear slab having length $L_S$ in the $z$-direction is placed in front of the left slit and completely covers it. The nonlinear dependence of the refractive index of the material is assumed to be of the Kerr type and is given by $n(I)=n_0+\gamma I$, where $I$ is the beam intensity, $n_0$ is the linear refractive index of the material and $\gamma$ is the Kerr coefficient. 
After the beams have propagated through the air and the nonlinear material, the intensity pattern is recorded at the observation plane $z=z_f$. 

In the paraxial approximation the beam dynamics inside the nonlinear material is given by
\begin{equation}
i\psi_z+\frac1{2k}\nabla_{x,y}^2\psi
+\gamma k_0|\psi|^2\psi=0
\label{eq:paraxial}
\end{equation}
where $\psi$ is the field amplitude, $\nabla_{x,y}^2=\partial_x^2+\partial_y^2$, $k=n_0\omega/c=n_0k_0$ is the wavenumber, $\omega$ is the optical frequency, and $c$ is the speed of light. When the beam propagates through the air $k=k_0$ and the nonlinearity is ignored in Eq.~(\ref{eq:paraxial}). At this point it is important to identify the relevant length scales of the problem. 
To this end we introduce normalized coordinates: 
We scale the field to the square root of the input intensity $\psi=I_0^{1/2}\Psi$. Furthermore, we measure distances in the transverse plane according to the aperture size along the $x$-direction ($x_0=w_x$), $X=x/x_0$, $Y=y/x_0$ and in the propagation direction according to the diffraction (Rayleigh) length $Z=z/z_0$, where $z_0=L_D=kx_0^2$. 
Therefore Eq.~(\ref{eq:paraxial}) can be written as 
\begin{equation}
i\Psi_Z +\frac12\nabla_{X,Y}^2\Psi+
\sgn(\gamma)\frac{L_D}{L_{NL}}|\Psi|^2\Psi=0,
\label{eq:paraxial_normalized}
\end{equation}
where the nonlinear length is defined as $L_{NL}=1/(\gamma I_0k_0)$. The length scales can be written in dimensionless form in terms of the diffraction length as $l_S=L_S/L_D$, $l_{NL}=L_{NL}/L_D$, $Z_f=z_f/L_D$. Finally, the observation plane ($z=z_f$) can be selected to be in the near (Fresnel) field or in the far field.

The relations between the relevant length scales of the nonlinear material play an important role in the dynamical behavior of the system. For weak beam intensities the nonlinear length is large. However, as the intensity increases the nonlinear length can take values that are much smaller than the diffraction length $L_{NL}\ll L_D$. We select a slab length and a slit aperture such that inside the slab the effect of diffraction is not dominant in the dynamics $L_S\ll L_D$. However, as the input intensity increases it is essential for the observation of nonlinear phenomena at the output that the nonlinear length and the slab length become of the same order of magnitude $L_{NL}\sim L_s$. Although a beam that passes from a square aperture diffracts quite fast due to the presence of discontinuities, we assume that the semiclassical, or ray optics, limit can be applied. This might look like a crude approximation however, as we are going to see, it captures surprisingly well the fundamental features of the diffracted pattern.

\begin{figure}[tb]
\centerline{\includegraphics[width=\columnwidth]{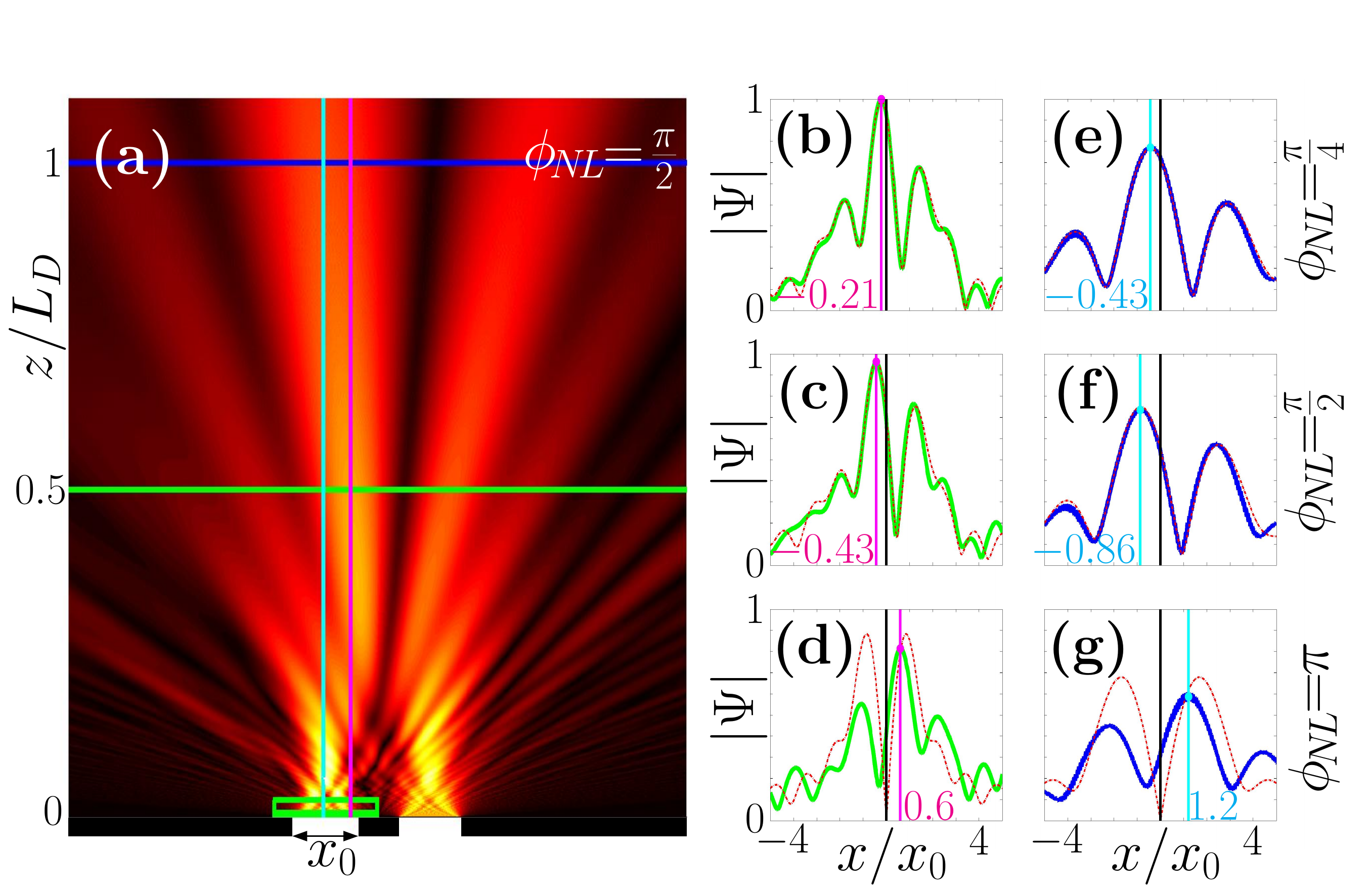}}
\caption{(color online) One-dimensional diffraction dynamics of a self-focusing nonlinear slab (green square) with length $L_s=L_D/40$ and aperture separation $w=5w_x/3$. In the three rows of (b)-(g) the field amplitude profile for phase accumulations $\phi_{NL}=\pi/4$, $\pi/2$, and $\pi$ is shown while in the two columns the observation plane is $z_f=0.5L_D$ and $z_f=L_D$, respectively. The solid curves are numerical results whereas the dashed curves are theoretical predictions using the Fresnel approximation. In all figures the transverse shift of the intensity maximum with respect to the origin is also shown with vertical lines.  In (a) the intensity dynamics for $\phi_{NL}=\pi$ is depicted and the horizontal lines correspond to the cross sections shown in (d) and (g).\label{fig:2}}
\end{figure}
Assuming that the apertures are equally and evenly illuminated leads to the following initial condition
\[
\psi(z=0)=
\sqrt{I_0}\left[\rect\frac{x+w/2}{w_x}+\rect\frac{x-w/2}{w_x}\right]
\rect\frac y{w_y}
\]
In the semiclassical limit, we can find a simplified expression for the field that passes through the slab: The amplitude remains invariant while the phase accumulates the nonlinear contribution
\begin{equation}
\phi_{NL}=\sgn(\gamma)\frac{L_S}{L_{NL}}
=\sgn(\gamma)\frac{l_S}{l_{NL}}.
\label{eq:phinl}
\end{equation}
In addition inside the slab the beam obtains a constant linear phase $\phi_L=n_0k_0L_S$. Since $\phi_L$ does not affect the dynamics as a function of the intensity through the rest of the paper for simplicity we assume that $n_0=1$. 
Note that through the rest of the paper we are going to use $\phi_{NL}$ not only as an estimate of the accumulated phase but, more importantly, as a measure of the relative scales between the nonlinear length $L_{NL}$ and the length of the slab $L_S$. Thus $\phi_{NL}$ increases by decreasing the normalized nonlinear length (i.e. increasing $\gamma$ or $I_0$) or by
increasing the normalized length of the slab. Because of this duality we prefer to present $\phi_{NL}$ in the figures instead of $L_{NL}$ and $L_S$. 
Asymptotically, as we described above, we can take into account the presence of the slab by modifying the initial condition to 
\[
\psi=\sqrt{I_0}\left[e^{i\phi_{NL}}\rect\frac{x+w/2}{w_x}+\rect\frac{x-w/2}{w_x}\right]
\rect\frac y{w_y}
\]
which then effectively propagates through the air for $z=z_f$. 
In the Fresnel limit the diffraction dynamics lead to
\begin{equation}
\psi=\sqrt{I_0}\left[
e^{i\Phi_{NL}}
{\cal I}(x,w/2,w_x)+
{\cal I}(x,-w/2,w_x)
\right] {\cal I}(y,0,w_y)
\label{eq:fresnel}
\end{equation}
where
\[
{\cal I}(x,\xi_c,w_x) = 
\frac{1}{\sqrt{2i}}[F(W_+(x,\xi_c,w_x))-F(W_-(x,\xi_c,w_x))],
\]
the function $F(t)$ is the following sum of the Fresnel integrals
$
F(t)=C(t)+iS(t) = \int_0^te^{i\pi s^2/2}ds,
$
and 
$
W_\pm(x,\xi_c,w_x) = \sqrt{k/(\pi z)}
\left(x+\xi_c\pm w_x/2\right).
$
The limiting case of 1D diffraction is obtained from the above equations by taking the limit $w_y\rightarrow\infty$ or ${\cal I}(y,0,w_y)\rightarrow1$.

\begin{figure}[tb]
\centerline{\includegraphics[width=\columnwidth]{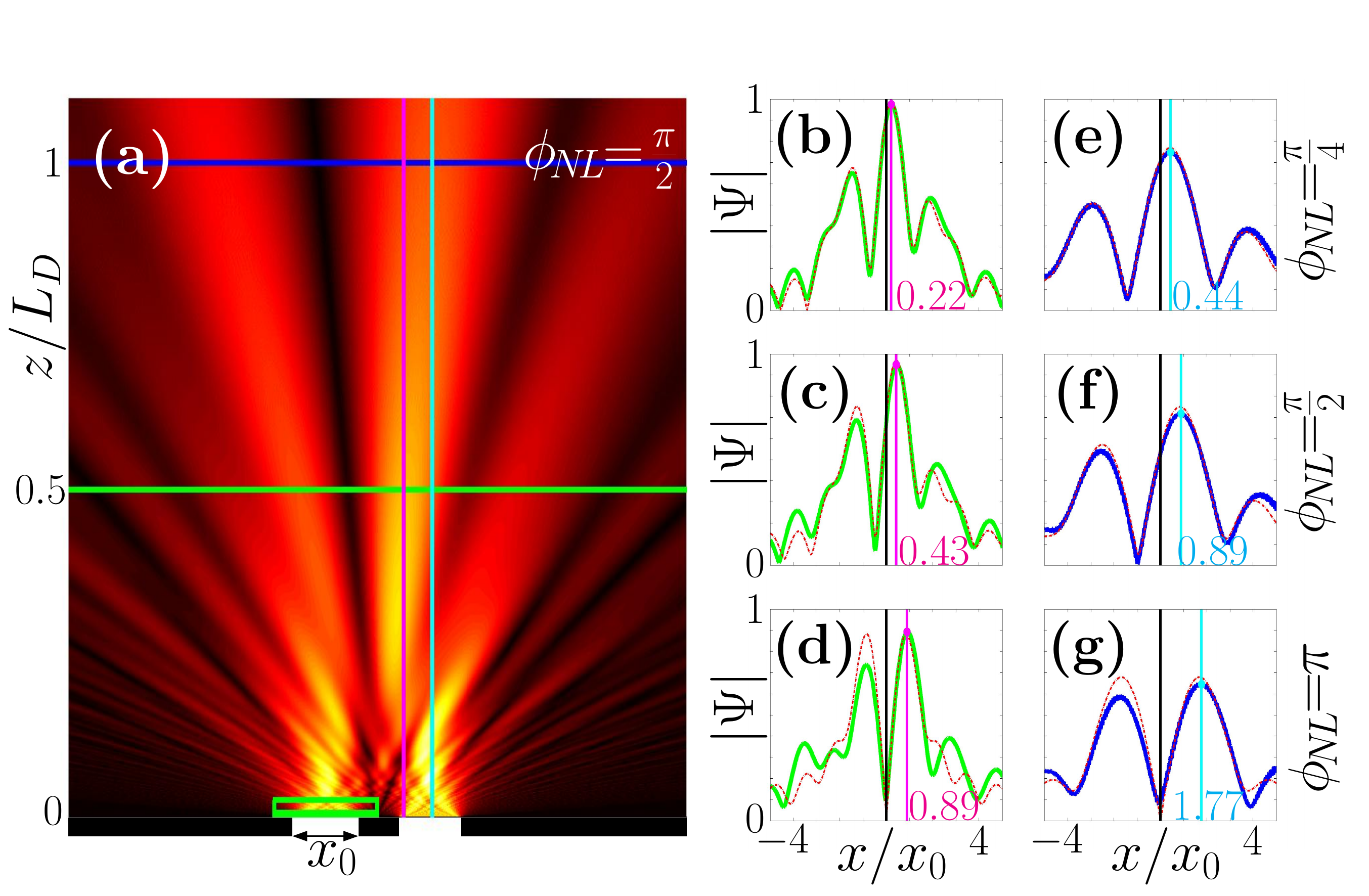}}
\caption{(color online) Same as in Fig.~\ref{fig:2} for self-defocusing nonlinearity. The values of $\phi_{NL}$ are the opposite as compared to Fig.~\ref{fig:2}. \label{fig:3}}
\end{figure}
\begin{figure}[tb]
\centerline{\includegraphics[width=\columnwidth]{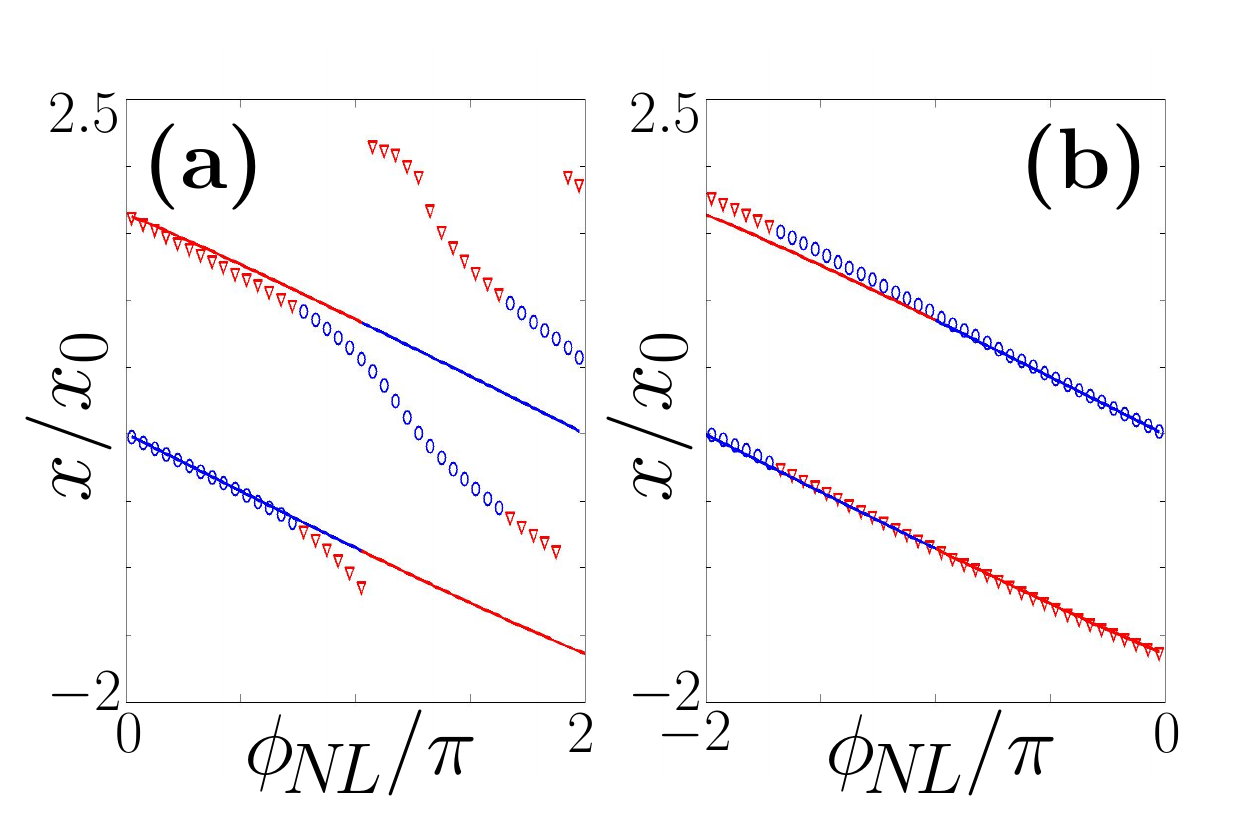}}
\caption{(color online) Locations of the two strongest intensity maxima as a function of the nonlinearity $\phi_{NL}$ for (a) self-focusing and (b) self-defocusing nonlinearity. The locations of the first and second maxima are shown with (blue) circles and (red) triangles, respectively, whereas the solid curves of the same colors are the theoretical predictions.  \label{fig:4}}
\end{figure}
In Fig.~\ref{fig:2} we see typical one-dimensional numerical results for the diffraction dynamics in the case of a slab of constant length with self-focusing nonlinearity in comparison with the asymptotic expression of Eq.~(\ref{eq:fresnel}). Our calculations are surprising accurate in the case where the nonlinear phase is $\phi_{NL}=\pi/4$ and $\phi_{NL}=\pi/2$ taking into account the simplicity of our assumptions as well as the relative large amount of diffraction inside the slab. Not only we can predict the location of the intensity maxima but also the amplitude profile of the main lobes. By further increasing the light intensity to $\phi_{NL}=\pi$ the numerical results start to slightly deviate from the predicted values.
The main reason is that the beam inside the slab, due to its higher intensity profile, starts to experience self-focusing. 
This behavior can be qualitatively understood in comparison with Fig.~\ref{fig:4}(a), where the location of the two first intensity maxima as a function of $\phi_{NL}$ for constant slab length is shown along with the theoretical predictions of Eq.~(\ref{eq:fresnel}). The beam passing from the left slab experiences self-focusing and its optical path length ($\mathrm{OPL}=\int n(s)ds$) increases with the intensity. As a result, the interference between the two slits is shifted almost linearly towards the negative $x$ direction. 
For $\phi_{NL}\lesssim\pi$ we see that the numerical data are slightly shifted in the negative $x$-direction as compared to the theoretical curve. This is the outcome of the weak self-focusing that further increases the accumulated phase (and thus the optical path) as compared to the predicted values.
As the nonlinearity increases, and for $\phi_{NL}\simeq0.74\pi$, the absolute maximum of the output intensity pattern is shifted to the adjacent maximum curve located to the right. The theoretical value for this jump is $\phi_{NL}=\pi$. We attribute this difference to the nonlinear spectral broadening of the beam passing through the slab that leads to increased diffraction, which eventually overcomes the amplitude increase due to self-focusing. Specifically, in Fig.~\ref{fig:2}(d), where $\phi=\pi$ we see that the left main lobe has significantly reduced intensity as compared to the theory, whereas the right main lobe compares well to the theoretical curve. By further increasing $\phi_{NL}$ up to $2\pi$ the same behavior is repeated: the intensity maximum is shifted to the next adjacent branch that is located to the right for $\phi_{NL}\simeq1.66\pi$, whereas the Fresnel theory predicts this value to be $\phi_{NL}=2\pi$. An even further increase of the nonlineariry $\phi_{NL}$ results to strong self-focusing inside the slab. The diffraction dynamics lead to more complicated intensity patterns at the observation plane that can not be predicted in terms of our theory.

\begin{figure}[tb]
\centerline{\includegraphics[width=\columnwidth]{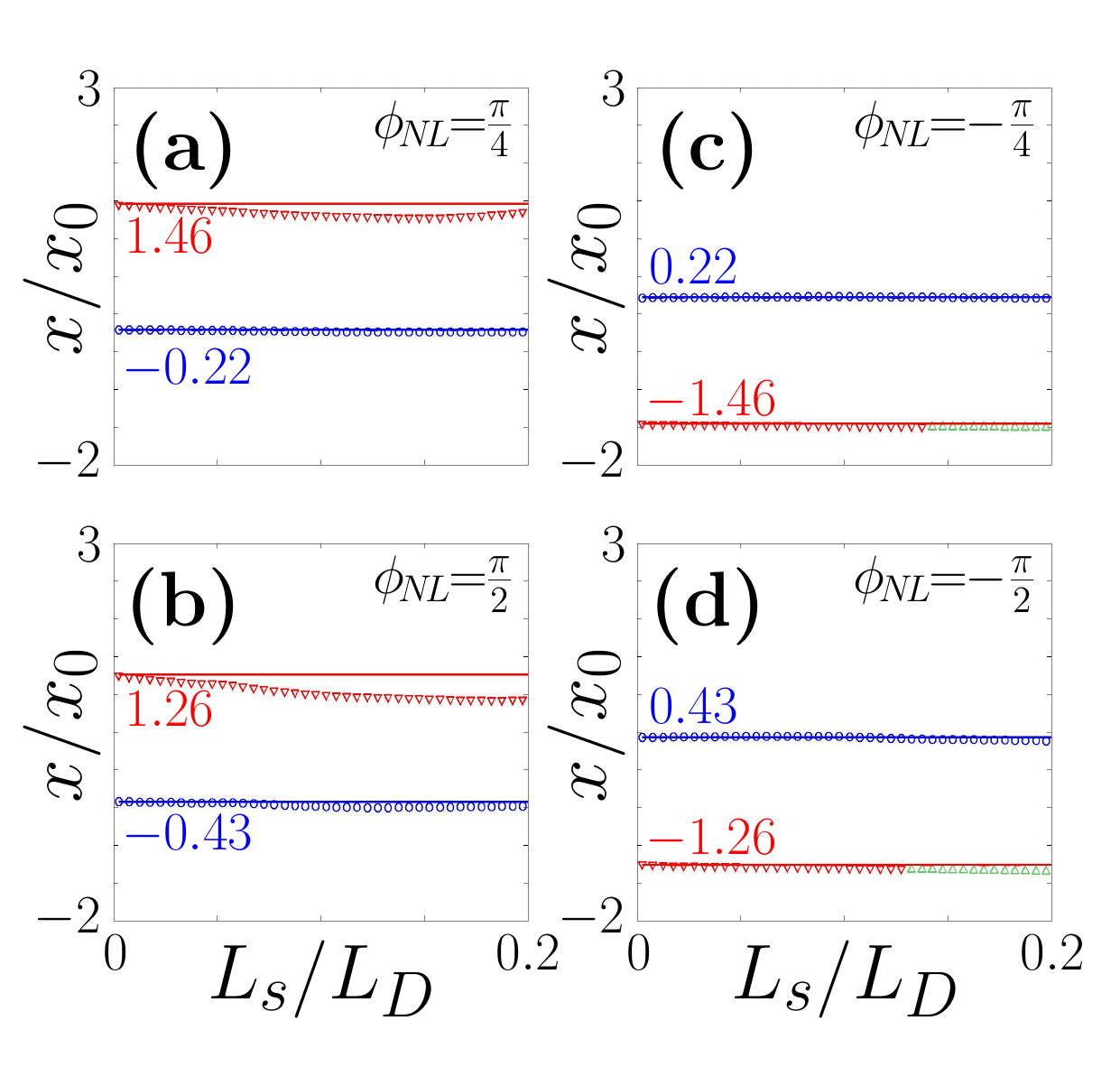}}
\caption{(color online) Locations of the first two intensity maxima as a function of the normalized slab length $l_S=L_S/L_D$ for constant phase accumulation $\phi_{NL}$. In the top and bottom rows $|\phi_{NL}|=\pi/4$ and $|\phi_{NL}|=\pi/2$. In the left and right columns the nonlinearity is self-focusing and self-defocusing respectively. The locations of the first and second maxima is shown with (blue) circles and (red) triangles, respectively, whereas the solid lines with the same colors are the predicted values.  In (d) the green triangles correspond to the
location of the third maximum.\label{fig:5}}
\end{figure}

The case of a nonlinear self-defocusing slab is shown in Fig.~\ref{fig:3}. As one can see, in this regime the theoretical and the numerical results are overall in even better agreement for larger values of the nonlinearity as compared to the self-focusing case. The defocusing nonlinearity reduces the effective index and thus the optical path length inside the nonlinear slab. As a result we expect that the intensity maxima, occurring due to constructive interference, are going to be shifted in the right direction. The agreement between theory and numerics for the two main lobes is very good even in the case where $\phi_{NL}=-2\pi$ [Fig.~\ref{fig:3}(d)-(g)]. 
In Fig.~\ref{fig:4}(b) we see the locations of the two strongest intensity maxima of the diffracted pattern in the self-defocusing regime as a function of the nonlinear parameter $\phi_{NL}$ for constant slab length. The agreement between theory and simulations is very good in this case for all the presented range of $\phi_{NL}$. The numerically obtained curves are slightly shifted to the right for large values of $|\phi_{NL}|$. This happens because, due to defocusing, the peak intensity gradually decreases inside the slab leading to a slightly reduced effective optical path as compared to the theory. Let us point out that the highest intensity peak remains in the same branch up until $|\phi_{NL}|\approx1.7\pi$, a value that is significantly larger than the theoretical prediction $|\phi_{NL}|=\pi$. This is due to the defocusing of the beam passing through the slab that decreases its intensity profile. As the absolute value of $\phi_{NL}$ is further increased the peak intensity translates to the left branch. 

The nonlinear two-slit setup shown in Fig.~\ref{fig:1} might be useful in nonlinear interferometric measurements. For example, it might be possible to measure the Kerr nonlinearity of a material from the shift in the resulting intensity pattern by utilizing the numerically obtained curves of Fig.~\ref{fig:4}.

\begin{figure}[tb]
\centerline{\includegraphics[width=\columnwidth]{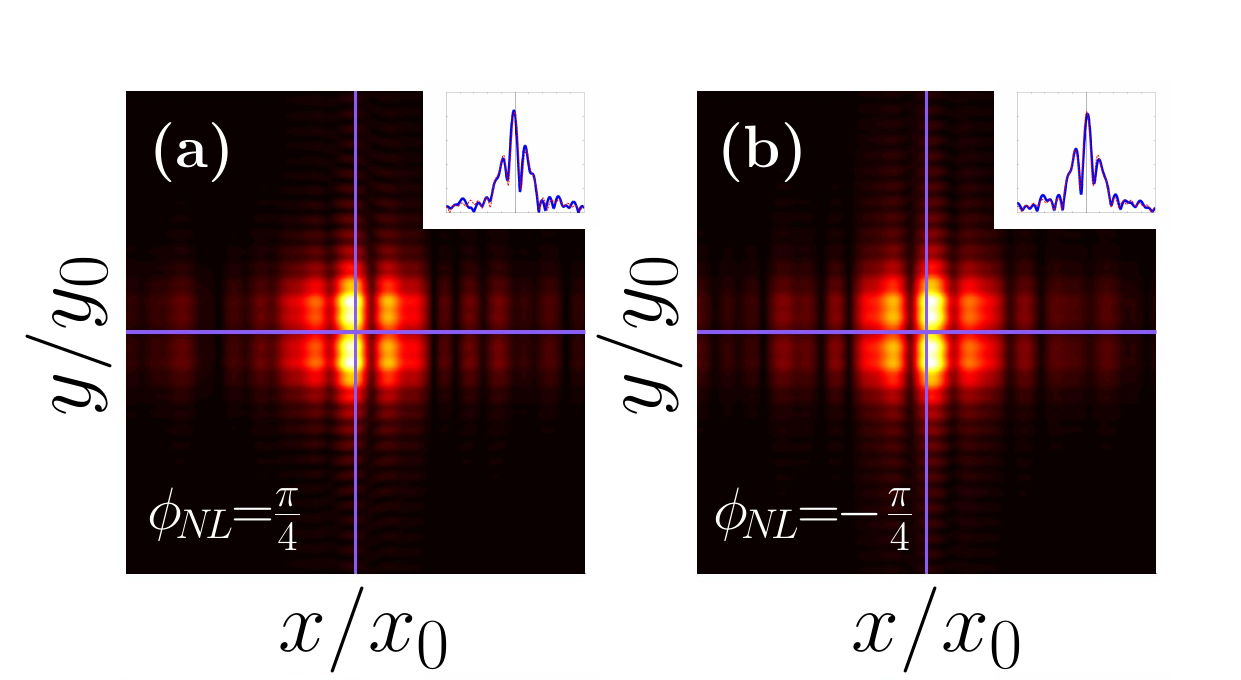}}
\caption{(color online) Two-dimensional diffraction dynamics in the case of self-focusing (a) and self-defocusing (b) nonlinearity for $|\phi_{NL}|=\pi/4$. The axes are shown with purple lines. In the insets cross sections at the $y=0$ plane are shown (solid blue) along with the predicted profile (dashed red) for $z=0.5L_D$. In all cases, $L_S/L_D=1/40$, each aperture has dimensions $w_x=(1,5)$, and aperture separation is $w=5/3$.\label{fig:6}}
\end{figure}
In Fig.~\ref{fig:5} we analyze the effect of the normalized slab length $l_S=L_S/L_D$ on the location of the intensity maxima for constant values of the total accumulated phase $\phi_{NL}=\sgn(\gamma)l_S/l_{NL}$. Thus by increasing the slab length $l_S$ we also need to proportionally increase the nonlinear length $l_{NL}$ to keep a constant value of $\phi_{NL}$. 
According to our model a constant phase accumulation $\phi_{NL}$ in slabs of different lengths is going to have the same effect. However, this is not in general true. In the limit where the slab length goes to zero, the effect of diffraction becomes negligible inside the slab and the Kerr effect results to purely self-phase modulation. Our model should be in perfect agreement with this limit. However, as the width of the slab increases, diffraction starts to become more and more important. The Kerr effect in combination with diffraction lead to the self-focusing or to the self-defocusing of the beam. In Fig.~\ref{fig:5} we see that for $|\phi_{NL}|=\pi/4,\pi/2$ and for slab lengths up to $1/5$ of the diffraction length we can accurately predict the location of the two first maxima for both signs of the nonlinearity. Small deviations occur in the case of self-focusing nonlinearity especially in the prediction of the location of the second intensity maximum. 

Finally, we have carried out numerical simulations of the proposed system in the case of two transverse directions. Typical results are shown in Fig.~\ref{fig:6} for $|\phi_{NL}|=\pi/4$. As one can see the agreement between the theory and the simulations is excellent. 

It might be interesting to present an example in physical units of the nonlinear double-slit configuration presented in this work. Specifically, for aperture size $x_0=100\,\mu m$, wavelength $\lambda=1\,\mu m$, index $n_0=2$ and slab size $l_S=L_S/L_D=1/40$ we obtain $L_D=63\,mm$ and $L_S=1.57\,mm$. A maximum phase accumulation $\phi_{NL}=\pi$ is obtained for index contrast $\Delta n_{NL}=\gamma I_0 = 1.59\times10^{-4}$.
Let us point out that an analytical expression can be obtained for the total power $P = 2\phi_{NL}/(n_0\gamma k_0^2l_S)$ which depends on the material and light properties as well as on $l_S$. For $\gamma=10^{-13}\,cm^2/W$ this results to a maximum double-slit laser power $P = 318\,kW$.

\rev{
In conclusion, we have studied a nonlinear interferometric setup that is based on diffractive interference of waves rather than beam combining. 
We expect that such a double-slit setup might be useful in nonlinear interferometry, for example in measuring the optical nonlinearities of materials. Of particular interest can be the possibility to explore the dynamics beyond the thin slab approximation where stronger nonlinearities can lead to more complicated dynamics. For example, strong self-focusing nonlinearities can lead to soliton generation. On the other hand, strong self-defocusing field exhibit increased nonlinear diffraction and amplitude decrease inside the slab. 
Of particular interest it to study the quantum limit of such a nonlinear double slit configuration. 
}

\newcommand{\noopsort[1]}{} \newcommand{\singleletter}[1]{#1}

\cleardoublepage

\renewcommand\refname{REFERENCES WITH TITLES}




\begin{thebibliography}{10}
\newcommand{\enquote}[1]{``#1''}

\bibitem{feynman-vol3-1963}
R.~Feynman, \emph{The Feynman lectures on physics: Vol 3, Quantum Mechanics}
  (Addison-Wesley Pub. Co, Reading, Mass, 1963). Chapter 1.

\bibitem{young-ptrs1804}
T.~Young, \enquote{Experimental demonstration of the general law of the
  interference of light,} Philos. T. R. Soc. Lond. \textbf{94}, 1 (1804).

\bibitem{taylo-pcps1909}
G.~Taylor, \enquote{Interference fringes with fibble light,} Proc. Camb.
  Philos. Soc. \textbf{15}, 114--115 (1909).

\bibitem{johns-zp1961}
C.~J\"onsson, \enquote{Elektroneninterferenzen an mehreren künstlich
  hergestellten feinspalten,} Zeitschrift f\"ur Physik \textbf{161}, 454--474
  (1961).

\bibitem{schou-prl2005}
H.~F. Schouten, N.~Kuzmin, G.~Dubois, T.~D. Visser, G.~Gbur, P.~F.~A. Alkemade,
  H.~Blok, G.~W.~t. Hooft, D.~Lenstra, and E.~R. Eliel,
  \enquote{Plasmon-assisted two-slit transmission: Young's experiment
  revisited,} Phys. Rev. Lett. \textbf{94}, 053901 (2005).

\bibitem{roman-oe2006}
J.~S. Roman, C.~Ruiz, J.~A. Perez, D.~Delgado, C.~Mendez, L.~Plaja, and
  L.~Roso, \enquote{Non-linear young's double-slit experiment,} Opt. Express
  \textbf{14}, 2817--2824 (2006).

\bibitem{sun-ls2008}
C.~Sun and J.~W. Fleischer, \enquote{Double slit diffraction in self-defocusing
  nonlinear media with nonlocal response,} in \enquote{Laser Science,}
  (Optical Society of America, 2008), p. JWA65.

\bibitem{gache-prl2010}
D.~Gachet, S.~Brustlein, and H.~Rigneault, \enquote{Revisiting the young's
  double slit experiment for background-free nonlinear raman spectroscopy and
  microscopy,} Phys. Rev. Lett. \textbf{104}, 213905 (2010).

\bibitem{dugua-apl1969}
M.~A. Duguay and J.~W. Hansen, \enquote{An ultrafast light gate,} Applied
  Physics Letters \textbf{15}, 192--194 (1969).

\bibitem{milbu-prl1989}
G.~J. Milburn, \enquote{Quantum optical fredkin gate,} Phys. Rev. Lett.
  \textbf{62}, 2124--2127 (1989).

\bibitem{agrawal-applications}
G.~P. Agrawal, \emph{Applications of Nonlinear Fiber Optics} (Academic Press,
  Burlington, 2008), 2nd ed.

\bibitem{sheik-ol1989}
M.~Sheik-bahae, A.~A. Said, and E.~W.~V. Stryland, \enquote{High-sensitivity,
  single-beam n2 measurements,} Opt. Lett. \textbf{14}, 955--957 (1989).

\end{thebibliography}
\end{document}